\documentclass[11pt]{article}

\usepackage{amssymb}     %
\usepackage{amsthm}      %
\usepackage[usenames]
           {color}       %
\usepackage{enumerate}   %
\usepackage{epsfig}      %
\usepackage{floatflt}    %
\usepackage{graphicx}    %
\usepackage{latexsym}    %
\usepackage{subfigure}   %

\newtheorem{lemma}{Lemma}[section]
\newtheorem{theorem}{Theorem}

\title{
  {\small University of Waterloo Technical Report CS-2007-14}\\
  An Approximation Algorithm for Shortest Descending Paths
}
\author{Mustaq Ahmed~~and~~Anna Lubiw \\
David R. Cheriton School of Computer Science \\
University of Waterloo,
Waterloo, ON, N2L 3G1, Canada \\
Email: \{\texttt{m6ahmed,alubiw}\}\texttt{@uwaterloo.ca}
}

\begin{document}

\pagenumbering{arabic} \pagestyle{plain}
\maketitle

\begin{abstract}
  A path from $s$ to $t$ on a polyhedral terrain is \emph{descending}
  if the height of a point $p$ never increases while we move $p$ along
  the path from $s$ to $t$. No efficient algorithm is known to find a
  shortest descending path (SDP) from $s$ to $t$ in a polyhedral
  terrain. We give a simple approximation algorithm that solves the
  SDP problem on general terrains. Our algorithm discretizes the
  terrain with $O( n^2 X / \epsilon )$ Steiner points so that after an
  \(
    O \left(
      \frac{n^2 X}{\epsilon} \log \left( \frac{n X}{\epsilon} \right)
    \right)
  \)
  -time preprocessing phase for a given vertex $s$, we can determine a
  $(1+\epsilon)$-approximate SDP from $s$ to any point $v$ in $O(n)$
  time if $v$ is either a vertex of the terrain or a Steiner point,
  and in $O( n X /\epsilon )$ time otherwise. Here $n$ is the size of
  the terrain, and $X$ is a parameter of the geometry of the terrain.
\end{abstract}

\section{Introduction}
\label{L2:Intro}

Finding a shortest path between two points in a geometric domain is
one of the fundamental problems in computational
geometry~\cite{Mitchell.98}. One extensively-studied version of the
problem is to compute a shortest path on a polyhedral terrain; this
has many applications in robotics, industrial automation, Geographic
Information Systems and wire routing. Our paper is about a variant of
this problem for which no efficient algorithm is known, the
\emph{Shortest Descending Path (SDP) Problem\/}: given a polyhedral
terrain, and points $s$ and $t$ on the surface, find a shortest path
on the surface from $s$ to $t$ such that, as a point travels along the
path, its elevation, or $z$-coordinate, never increases. We need to
compute a shortest descending path, for example, for laying a canal of
minimum length from the source of water at the top of a mountain to
fields for irrigation purpose~\cite{Roy.05}, and for skiing down a
mountain along a shortest route.

The SDP problem was introduced by De Berg and van
Kreveld~\cite{Berg.97}, who gave an algorithm to preprocess a terrain
in $O(n\log n)$ time so that it can be decided in $O(\log n)$ time if
there exists a descending path between any pair of vertices. They did
not consider the length of the path, and left open the problem of
finding the shortest such path. Roy, Das and Nandy~\cite{Roy.05}
solved the SDP problem for two special classes of terrains. For convex
(or concave) terrains, they use the continuous Dijkstra approach to
preprocess the terrain in $O(n^2 \log n)$ time and $O(n^2)$ space so
that an SDP of size $k$ can be determined in $O(k + \log n)$ time. For
a terrain consisting of edges parallel to one another, they find an
SDP in $O(n \log n)$ time by transforming the faces of the terrain in
a way that makes the unfolded SDP a straight line segment. In our
previous paper~\cite{Ahmed.06} we examined some properties of SDPs,
and gave an $O(n^{3.5} \log(\frac{1}{\epsilon}))$ time
$(1+\epsilon)$-approximation algorithm that finds an SDP through a
\emph{given\/} sequence of faces, by formulating the problem as a
convex optimization problem.

In this paper we present a $(1+\epsilon)$-approximation algorithm,
which is the first algorithm to solve the SDP problem on general
terrains. Given a vertex $s$ in a triangulated terrain, and a constant
$\epsilon \in (0,1]$, we discretize the terrain with
\(
  O \left( \frac{n^2 X}{\epsilon} \right)
\)
Steiner points so that after an
\(
  O \left( n^3 \left(\frac{X}{\epsilon}\right)^2 \right)
\)
-time preprocessing phase for a given vertex $s$, we can determine a
$(1+\epsilon)$-approximate SDP from $s$ to any point $v$ in $O(n)$
time if $v$ is either a vertex of the terrain or a Steiner point, and
in
\(
  O \left( \frac{n X}{\epsilon} \right)
\)
time otherwise, where $n$ is the number of vertices of the terrain,
and $X$ is a parameter of the geometry of the terrain. More precisely,
$X = \frac{L}{h} \cdot \frac{1}{\cos\theta} = \frac{L}{h} \sec\theta$,
where $L$ is the length of the longest edge, $h$ is the smallest
distance of a vertex from a non-adjacent edge in the same face, and
$\theta$ is the largest acute angle between a non-level edge and a
perpendicular line. We achieve this result by discretizing the terrain
with Steiner points along the edges---the main trick is to ensure the
\emph{existence\/} of a descending path through the Steiner points
that approximates the SDP. The algorithm is very simple, and hence
easy to implement.

The paper is organized as follows. We define a few terms and discuss
the terrain parameter $X$ in Sect.~\ref{L3:Prelim}, and then mention
related results in Sect.~\ref{L3:Background} and~\ref{L3:Bushwhack}.
Section~\ref{L2:SteinerPt} gives the details of our approximation
algorithm. In Sect.~\ref{L2:Conclusion} we mention our ongoing work,
and discuss the possibility of an exact solution using the approach of
Chen and Han~\cite{Chen.96}.

\subsection{Preliminaries}
\label{L3:Prelim}

A terrain is a 2D surface in 3D space with the property that every
vertical line intersects it in a point~\cite{Berg.00}. For any point
$p$ in the terrain, $h(p)$ denotes the height of $p$, i.e.,~the
$z$-coordinate of $p$. An \emph{isoline\/} on a non-level face is a
line through two points of equal height on that face. We add $s$ as a
vertex of the terrain and triangulate the terrain in $O(n)$
time~\cite{Chazelle.91}. Since $n$ is the number of vertices in the
terrain, it follows from \emph{Euler's
formula\/}~\cite[Page~29]{Berg.00} that the terrain has at most $3 n$
edges, and at most $2 n$ faces.
  
A path $P$ from $s$ to $t$ on the terrain is \emph{descending\/} if
the $z$-coordinate of a point $p$ never increases while we move $p$
along the path from $s$ to $t$. A line segment of a descending path in
face $f$ is called a \emph{free segment\/} if moving either of its
endpoints by an arbitrarily small amount to a new position in $f$
keeps the segment descending. Otherwise, the segment is called a
\emph{constrained segment}. All the points in a constrained segment
are at the same height, though not all constant height segments are
constrained. For example, a segment in a level face is free, although
all its points are at the same height. Clearly, a constrained segment
can only appear in a non-level face, and it is an isoline in that
face. A path consisting solely of free [constrained] segments is
called a \emph{free path\/} [\emph{constrained path}, respectively].

\begin{figure}[t]
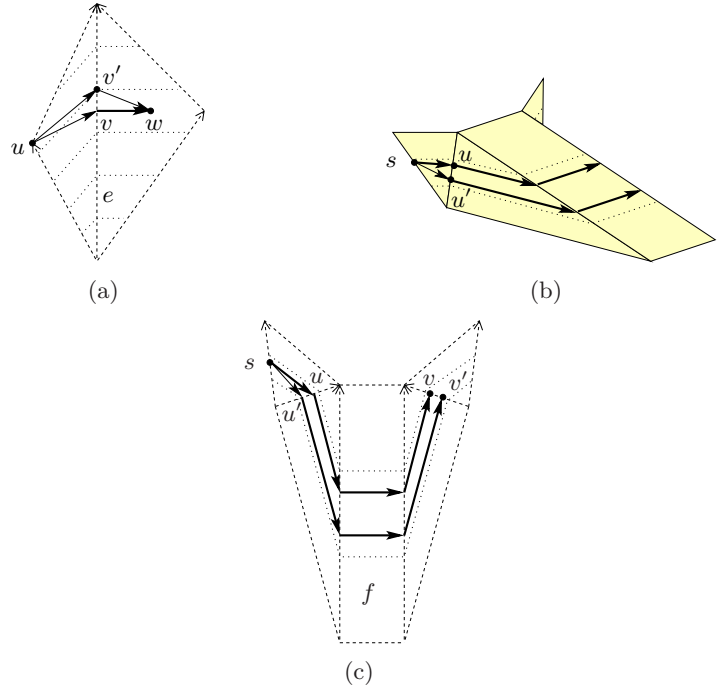

  \hspace*{\fill}
  \subfigure[]{
    \input{figSkinnyTriangles.pstex_t}
    \label{fig:SkinnyTriangles}
  }
  \hspace*{\fill}
  \subfigure[]{
    \input{figStPtsExtreme1.pstex_t}
    \label{fig:StPtsExtreme1}
  }
  \hspace*{\fill}

  \hspace*{\fill}
  \subfigure[]{
    \input{figStPtsExtreme2.pstex_t}
    \label{fig:StPtsExtreme2}
  }
  \hspace*{\fill}

  \caption{The effect of $\frac{L}{h}$ and $\theta$ on approximate
    SDPs}
  \label{fig:GeometricEffect}
\end{figure}

We assume that all paths in our discussion are directed. Our
discussion relies on the following properties of an
SDP~\cite{Ahmed.06}: a subpath of an SDP is an SDP; and an SDP
intersects a face at most once. Note that an unfolded SDP is not
always a straight line segment, see Figure~\ref{fig:GeometricEffect}.

We use the term ``edge'' to denote a line segment of the terrain,
``vertex'' to denote an endpoint of an edge, ``segment'' to denote a
line segment of a path and ``node'' to denote an endpoint of a
segment. We use ``node'' and ``link'' to mean the corresponding
entities in a graph or a tree. In our figures, we use dashed lines for
edges, possibly marking the upward direction with arrows. A solid
arrow denotes a path segment, which may be heavy to mark a constrained
segment. Dotted lines are used to show the isolines in a face.

We will now discuss the two geometric parameters $\frac{L}{h}$ and
$\theta$. The first parameter $\frac{L}{h}$ is a 2D parameter, and
captures how ``skinny'' the terrain faces are. A terrain with a large
value of $\frac{L}{h}$ needs more Steiner points to approximate an
SDP, which is evident from the example in
Fig.~\ref{fig:SkinnyTriangles}. In this figure, $(u, v, w)$ is an SDP,
and $v'$ is the nearest Steiner point on edge $e$. By making $u$ and
$w$ very close to $e$, and thus making $\frac{L}{h}$ large, the ratio
of the lengths of the paths $(u, v', w)$ and $(u, v, w)$ can be made
arbitrarily large, which necessitates more Steiner points on $e$ to
maintain a desired approximation factor. Such effects of skinny
triangles are well known~\cite{Berg.00}, and have been observed in the
Steiner point approaches for other shortest path problems,
e.g.,~Aleksandrov et al.~\cite{Aleksandrov.05}.

The second parameter $\theta$ captures the orientation of the terrain
faces in 3D space. When $\theta$ is close to $\frac{\pi}{2}$ radians,
which means that there is an \emph{almost\/} level edge $e$, it is
possible to construct a pair of SDPs from $s$ that have their ending
nodes very close to each other, but cross $e$ at points that are far
apart. Figure~\ref{fig:StPtsExtreme1} shows two such SDPs on a terrain
that is shown unfolded in Fig.~\ref{fig:StPtsExtreme2} (The terrain
can be simplified though it becomes less intuitive). It can be shown
that both the paths are SDPs. Assuming that $u'$ is the closest
Steiner point from $u$ below $h(u) = h(s)$, the best \emph{feasible\/}
approximation of the path from $s$ to $v$ is the path from $s$ to
$v'$. The approximation factor can be made arbitrarily large by making
face $f$ close to level position, and thus making $\theta$ close to
$\frac{\pi}{2}$ radians, no matter how small $|u u'|$ is. Note that
the ``side triangles'' become skinny, which can be avoided by making
the two edges through $u u'$ and $v v'$ longer by moving their lower
vertices further down along the lines $u u'$ and $v v'$ respectively.

\subsection{Related Work}
\label{L3:Background}

It was Papadimitriou~\cite{Papadimitriou.85} who first introduced the
idea of discretizing space by adding Steiner points and approximating
a shortest path through the space by a shortest path in the graph of
Steiner points. He did this to find a shortest obstacle-avoiding path
in 3D---a problem for which computing an exact solution is
NP-hard~\cite{Canny.87}. On polyhedral surfaces, the Steiner point
approach has been used in approximation algorithms for many variants
of the shortest path problem, particularly those in which the shortest
path does not unfold to a straight line segment. One such variant is
the Weighted Region Problem~\cite{Mitchell.91}. In this problem, a set
of constant weights is used to model the difference in costs of travel
in different regions on the surface, and the goal is to minimize the
weighted length of a path. Mitchell and
Papadimitriou~\cite{Mitchell.91} used the continuous Dijkstra approach
to get an approximate solution in
\(
  O\left(
    n^8 \log\left(\frac{n}{\epsilon}\right)
  \right)
\)
time. Following their result, several faster approximation
schemes~\cite{Aleksandrov.98,Aleksandrov.00,Aleksandrov.05,Sun.06a}
have been devised, all using the Steiner point approach. The Steiner
points are placed along the edges of the terrain, except that
Aleksandrov et al.~\cite{Aleksandrov.05} place them along the
bisectors of the face angles. A comparison between these algorithms
can be found in Aleksandrov et al.~\cite{Aleksandrov.05}.

One generalization of the Weighted Region Problem is finding a
shortest aniso-tropic path~\cite{Rowe.90}, where the weight assigned
to a region depends on the direction of travel. The weights in this
problem capture, for example, the effect the gravity and friction on a
vehicle moving on a slope. Lanthier et al.~\cite{Lanthier.99}, Sun and
Reif~\cite{Sun.05} and Sun and Bu~\cite{Sun.06b} solved this problem
by placing Steiner points along the edges.

Note that all the above-mentioned Steiner point approaches place the
Steiner points in a face without considering the Steiner points in the
neighboring faces. This strategy works because we can travel in a face
in \emph{any\/} direction. For the shortest anisotropic path problem,
traveling in a ``forbidden'' direction within a face is possible by
following a zig-zag path. For the SDP problem, traveling in an
ascending direction is impossible---a fact that makes it a non-trivial
work to place the Steiner points.

\subsection{The Bushwhack Algorithm}
\label{L3:Bushwhack}

Our algorithm uses a variant of Dijkstra's algorithm, called the
Bushwhack algorithm~\cite{Sun.01}, to compute a shortest path in the
graph of Steiner points in a terrain. The Bushwhack algorithm achieves
$O(|V| \log |V|)$ running time by utilizing certain geometric
properties of the paths in such a graph. The algorithm has been used
in shortest path algorithms for the Weighted Region
Problem~\cite{Aleksandrov.05,Sun.06a} and the Shortest Anisotropic
Path problem~\cite{Sun.05}.

\begin{figure}[tb]
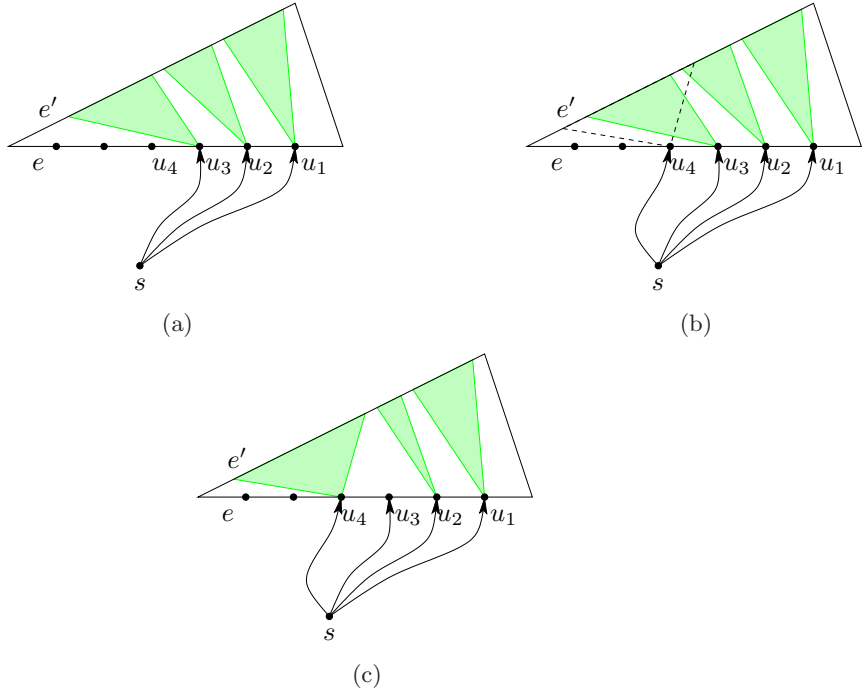

  \hspace*{\fill}
  \subfigure[]{
    \input{figOthers-Sun06-1.pstex_t}
    \label{fig:Others-Sun06-1}
  }
  \hspace*{\fill}
  \subfigure[]{
    \input{figOthers-Sun06-2.pstex_t}
    \label{fig:Others-Sun06-2}
  }
  \hspace*{\fill}
  \subfigure[]{
    \input{figOthers-Sun06-3.pstex_t}
    \label{fig:Others-Sun06-3}
  }
  \hspace*{\fill}
  \caption{Maintaining the list $I_{e,e'}$ in the Bushwhack algorithm}
  \label{fig:Others-Sun06}
\end{figure}

The Buskwhack algorithm relies on a simple, yet important, property of
shortest paths on terrains: two shortest paths through different face
sequences do not intersect each other at an interior point of a
face. As a result, for any two consecutive Steiner points $u_1$ and
$u_2$ on edge $e$ for which the distances from $s$ are already known,
the corresponding sets of ``possible next nodes on the path'' are
disjoint, as shown using shading in Figure~\ref{fig:Others-Sun06-1}.
This property makes it possible to consider only a subset of links at
a Steiner point $v$ when expanding the shortest path tree onwards from
$v$ using Dijkstra's algorithm. More precisely, Sun and Reif maintain
a dynamic list of intervals $I_{e,e'}$ for every pair of edges $e$ and
$e'$ of a common face. Each point in an interval is reachable from $s$
using a shortest path through a common sequence of intermediate
points. For every Steiner point $v$ in $e$ with known distance from
$s$, $I_{e,e'}$ contains an interval of Steiner points on $e'$ that
are likely to become the next node in the path from $s$ through
$v$. The intervals in $I_{e,e'}$ are ordered in accordance with the
ordering of the Steiner points $v$ on $e$, which enables easy
insertion of the interval for a Steiner point on $e$ whose distance
from $s$ is yet unknown. For example, right after the distance of
$u_4$ from $s$ becomes known (i.e.,~right after $u_4$ gets dequeued in
Dijkstra's algorithm) as shown in Figure~\ref{fig:Others-Sun06-2}, the
Steiner points on $e'$ that are closer to $u_4$ than to any other
Steiner points on $e$ with known distances from $s$ can be located in
time logarithmic in the number of Steiner points on $e'$, using binary
searches (Figure~\ref{fig:Others-Sun06-3}). Within the interval for
each Steiner point $u \in e$, only the Steiner point that is the
nearest one from $u$ is enqueued. Since the nearest Steiner point from
$u$ in its interval can be determined in constant time, each iteration
of the modified Dijkstra's algorithm (i.e.,~the Bushwhack algorithm)
takes $O(|V|)$ time, resulting in a total running time of $O(|V| \log
|V|)$.

\section{Approximation using Steiner Points}
\label{L2:SteinerPt}

Our approximation algorithm works by first discretizing the terrain
with many Steiner points along the edges, and then determining a
shortest path in a directed graph in which each directed link connects
a pair of vertices or Steiner points in a face of the terrain in the
descending (more accurately, in the non-ascending) direction. Because
of the nature of our problem, we determine the positions of the
Steiner points in a way completely different from the Steiner point
approaches discussed in Sect.~\ref{L3:Background}. In particular, we
cannot place Steiner points in an edge without considering the heights
of the Steiner points in other edges. We will now elaborate on this
issue before going through the details of our algorithm.

\subsection{Placing the Steiner Points}
\label{L3:SteinerPt.Placing}

\begin{figure}[tb]
  \hspace*{\fill}
  \input{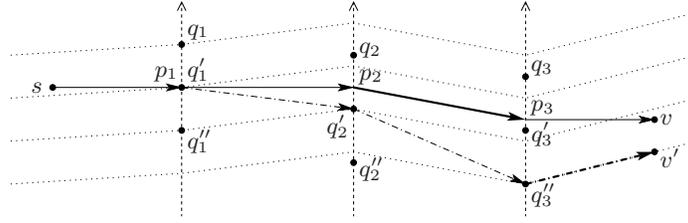}
  \hspace*{\fill}
  \caption{Problems with independently-placed Steiner points}
  \label{fig:IndependentStPts}
\end{figure}

For each Steiner point $p$ in an edge, if there is no Steiner point
with height $h(p)$ in other edges of the neighboring faces, it is
possible that a descending path from $s$ to $v$ through Steiner points
does not exist, or is arbitrarily longer than the SDP. For example,
consider the SDP $P = (s, p_1, p_2, p_3, v)$ in
Fig.~\ref{fig:IndependentStPts}, where for each $i \in [1,3]$, $q_i$,
$q'_i$ and $q''_i$ are three consecutive Steiner points with $h(q_i) >
h(q'_i) > h(q''_i)$ such that $q_i$ is the nearest Steiner point above
$p_i$. Note that $p_1$ and $q'_1$ are the same point in the
figure. There is no descending path from $s$ to $v$ through the
Steiner points: we must cross the first edge at $q'_1$ or lower, then
cross the second edge at $q'_2$ or lower, and cross the third edge at
$q''_3$ or lower, which puts us at a height below $h(v)$. Another
important observation is that even if a descending path exists, it may
not be a good approximation of $P$. In
Fig.~\ref{fig:IndependentStPts}, for example, if we want to reach
instead a point $v'$ slightly below $v$, $P'$ would be a feasible
path, but the last intermediate nodes of $P$ and $P'$ are not very
close. We can easily extend this example to an SDP $P$ going through
many edges such that the ``nearest'' descending path $P'$ gets further
away from $P$ at each step, and at one point, $P'$ starts following a
completely different sequence of edges. Clearly, we cannot ensure a
good approximation by just making the Steiner points on an edge close
to each other.

To guarantee the existence of a descending path through Steiner points
that approximates an SDP from $s$ to any vertex, we have to be able to
go through the Steiner points in a sequence of faces without ``losing
height'', i.e.,~along a constrained path. We achieve this by slicing
the terrain with a set of horizontal planes, and then putting Steiner
points where the planes intersect the edges. The set of horizontal
planes includes one plane through each vertex of the terrain, and
other planes in between them so that two consecutive planes are within
distance $\delta$ of each other, where $\delta$ is a small constant
that depends on the approximation factor.

One important observation is that our scheme makes the distance
between consecutive Steiner points on an edge dependent on the slope
of that edge. For instance, the distance between consecutive Steiner
points is more for an almost-level edge than for an almost vertical
edge. Since $\theta$ is the largest acute angle between a non-level
edge and a perpendicular line, it follows easily that the distance
between consecutive Steiner points on a non-level edge is at most
$\delta \sec\theta$. Because of the situation depicted in
Fig.~\ref{fig:IndependentStPts}, we cannot place extra Steiner points
\emph{only\/} on the edges that are almost level. Contrarily, we can
put Steiner points on a level edge without considering heights, since
a level edge can never result in such a situation (because all the
points in such an edge have the same height).

\subsection{Approximation Algorithm}
\label{L3:SteinerPt.Alg}

Our algorithm has two phases. In the preprocessing phase, we place the
Steiner points, and then construct a shortest path tree in the
corresponding graph. During the query phase, the shortest path tree
gives an approximate SDP in a straightforward manner.

\paragraph{Preprocessing phase.}
Let $\delta = \frac{\epsilon h \cos\theta}{4 n}$. We subdivide every
non-level edge $e$ of the terrain by putting Steiner points at the
points where $e$ intersects each of the following planes: $z = j
\delta$ for all positive integers $j$, and $z = h(x)$ for all vertices
$x$ of the terrain. We subdivide every level edge $e$ by putting
enough Steiner points so that the length of each part of $e$ is at
most $\delta \sec\theta$. Let $V$ be the set of all vertices and all
Steiner points in the terrain. We then construct a weighted graph
$G=(V,E)$ as follows, starting with $E = \emptyset$. For every pair
$(x, y)$ of points in $V$ adjacent to a face $f$ of the terrain, we
add to $E$ a directed link from $x$ to $y$ if and only if $h(x) \ge
h(y)$ and $x y$ is either an edge of the terrain or a segment through
the interior of $f$. Note that we do \emph{not\/} add a link between
two points on the same edge unless both of them are vertices. Each
link in $E$ is assigned a weight equal to the length of the
corresponding line segment in the terrain. Finally we construct a
shortest path tree $T$ rooted at $s$ in $G$ using the Bushwhack
algorithm.

Note that we are mentioning set $E$ only to make the discussion easy.
In practice, we do not construct $E$ explicitly because the neighbors
of a node $x \in V$ in the graph is determined \emph{during\/} the
execution of the Bushwhack algorithm.

\paragraph{Query phase.}
When the query point $v$ is a node of $G$, we return the path from $s$
to $v$ in $T$ as an approximate SDP. Otherwise, we locate the node $u$
among those in $V$ lying in the face(s) containing $v$ such that $h(u)
\ge h(v)$, and the sum of the length of the path from $s$ to $u$ in
$T$ and the length of the segment $u v$ is minimum. We return the
corresponding path from $s$ to $v$ as an approximate SDP.

\subsection{Correctness and Analysis}
\label{L3:SteinerPt.Analysis}

For the proof of correctness, it is sufficient to show that an SDP $P$
from $s$ to any point $v$ in the terrain is approximated by a
descending path $P'$ such that all the segments, except the last one,
of $P'$ exist in $G$. We show this by constructing a path $P'$ from
$P$ in the following way. Note that $P'$ might not be the path
returned by our algorithm, but it provides an upper bound on the
length of the returned path.

Let $P = (s=p_0, p_1, p_2, \ldots, p_k, v=p_{k+1})$ be an SDP from $s$
to $v$ such that $p_i$ and $p_{i+1}$ are two different boundary points
of a common face for all $i \in [0,k-1]$, and $p_k$ and $p_{k+1}$ are
two points of a common face. For ease of discussion, let $e_i$ be an
edge of the terrain through $p_i$ for all $i \in [1,k]$ ($e_i$ can be
any edge through $p_i$ if $p_i$ is a vertex). Intuitively, we
construct $P'$ by moving all the intermediate nodes of $P$ upward to
the nearest Steiner point. More precisely, we define a path $P' =
(s=p'_0, p'_1, p'_2, \ldots, p'_k, v=p'_{k+1})$ as follows. For each
$i \in [1,k]$, let $p'_i = p_i$ if $p_i$ is a vertex of the
terrain. Otherwise, let $p'_i$ be the nearest point from $p_i$ in $V
\cap e_i$ such that $h(p'_i) \ge h(p_i)$. Such a point always exists
in $V$ because $p_i$ is an interior point of $e_i$ in this case, and
it has two neighbors $x$ and $y$ in $V \cap e_i$ such that $h(x) \ge
h(p_i) \ge h(y)$.

\begin{lemma} \label{lem:Steiner.Feasibility}
  Path $P'$ is descending, and the part of $P'$ from $s$ to $p'_k$
  exists in $G$.
\end{lemma}

\begin{proof}
  We prove that $P'$ is descending by showing that $h(p'_i) \ge
  h(p'_{i+1})$ for every $i \in [0,k]$. We have: $h(p'_i) \ge
  h(p_{i+1})$, because $h(p'_i) \ge h(p_i)$ by the definition of
  $p'_i$, and $h(p_i) \ge h(p_{i+1})$ as $P$ is descending. Now
  consider the following two cases:
  \begin{description}
  \item[Case 1:] $p'_{i+1} = p_{i+1}$ or $e_{i+1}$ is a level edge.
    In this case, $h(p'_{i+1}) = h(p_{i+1})$. It follows from the
    inequality $h(p'_i) \ge h(p_{i+1})$ that $h(p'_i) \ge
    h(p'_{i+1})$.

  \item[Case 2:] $p'_{i+1} \neq p_{i+1}$ and $e_{i+1}$ is a non-level
    edge. In this case, there is either one or no point in $e_{i+1}$
    at any particular height. Let $p''_{i+1}$ be the point in
    $e_{i+1}$ such that $h(p''_{i+1}) = h(p'_i)$, or if no such point
    exists, let $p''_{i+1}$ be the upper vertex of $e_{i+1}$. In the
    latter case, we can infer from the inequality $h(p'_i) \ge
    h(p_{i+1})$ that $h(p'_i) > h(p''_{i+1})$. Therefore we have
    $h(p'_i) \ge h(p''_{i+1})$ in both cases. Since $p''_{i+1} \in V
    \cap e_{i+1}$, the definition of $p'_{i+1}$ implies that
    $h(p''_{i+1}) \ge h(p'_{i+1})$. So, $h(p'_i) \ge h(p'_{i+1})$.

  \end{description}
  Therefore, $P'$ is a descending path.

  To show that the part of $P'$ from $s$ to $p'_k$ exists in $G$, it
  is sufficient to prove that $p'_i p'_{i+1} \in E$ for all $i \in
  [0,k-1]$, because both $p'_i$ and $p'_{i+1}$ are in $V$ by
  definition. We have already proved that $h(p'_i) \ge
  h(p'_{i+1})$. Since $p'_i$ and $p'_{i+1}$ are boundary points of a
  common face by definition, $p'_i p'_{i+1} \not\in E$ only in the
  case that both of $p'_i$ and $p'_{i+1}$ lie on a common edge, and at
  most one of them is a vertex. We show as follows that this is
  impossible. When both $p_i$ and $p_{i+1}$ are vertices of the
  terrain, both $p'_i$ and $p'_{i+1}$ are vertices. When at least one
  of $p_i$ and $p_{i+1}$ is an interior point of an edge, they cannot
  lie on a common edge~\cite[Lemma~3]{Ahmed.06}; therefore, both of
  $p'_i$ and $p'_{i+1}$ cannot lie on a common edge unless both of
  $p'_i$ and $p'_{i+1}$ are vertices. So, this is impossible that both
  $p'_i$ and $p'_{i+1}$ lie on a common edge, and at most one of them
  is a vertex. Therefore, $p'_i p'_{i+1} \in E$.
\end{proof}
\begin{lemma} \label{lem:Steiner.ApproxFactor}
  Path $P'$ is a $(1+\epsilon)$-approximation of $P$.
\end{lemma}

\begin{proof}
  We first show that $\sum_{i=1}^k |p_i p'_i| < \frac{\epsilon
  h}{2}$. When $p_i \neq p'_i$, and $e_i$ is a non-level edge, we
  have: $|h(p_i) - h(p'_i)| \le \delta$ by construction, and
  $\frac{|h(p_i) - h(p'_i)|}{|p_i p'_i|} \ge \cos\theta$ using
  elementary trigonometry, which implies that $|p_i p'_i| \le \delta
  \sec\theta$. When $p_i \neq p'_i$, and $e_i$ is a level edge, $|p_i
  p'_i| \le \delta \sec\theta$ by construction. When $p_i = p'_i$,
  $|p_i p'_i| = 0$. Therefore, $\sum_{i=1}^k |p_i p'_i| \le k \delta
  \sec\theta$. Because the number of faces in the terrain is at most
  $2 n$, $k \le 2 n - 1$, and hence, $\sum_{i=1}^k |p_i p'_i| < 2 n
  \delta \sec\theta = \frac{\epsilon h}{2}$.

  Now, the length of $P'$ is equal to:
  \begin{eqnarray*}
    \sum_{i=0}^k |p'_i p'_{i+1}|
    &\le&
    \sum_{i=0}^k \left(
      |p'_i p_i| + |p_i p_{i+1}| + |p_{i+1} p'_{i+1}|
    \right)
    \\
    &=&
    \;<\;
    \sum_{i=0}^k |p_i p_{i+1}| + \epsilon h
    \enspace.
  \end{eqnarray*}
  Assuming that $P$ crosses at least one edge of the terrain
  (otherwise, $P' = (s,v) = P$), $\sum_{i=0}^k |p_i p_{i+1}| \ge h$,
  and therefore,
  \[
    \sum_{i=0}^k |p'_i p'_{i+1}|
    <
    (1+\epsilon) \sum_{i=0}^k |p_i p_{i+1}|
    \enspace.
  \]
  Because $P'$ is descending (Lemma~\ref{lem:Steiner.Feasibility}), it
  follows that $P'$ is a $(1+\epsilon)$-approximation of $P$.
\end{proof}
\begin{theorem} \label{thm:Steiner.ApproxAlg}
  Let $X = \left(\frac{L}{h}\right) \sec\theta$. Given a vertex $s$ in
  the terrain, and a constant $\epsilon \in (0,1]$, we can discretize
  the terrain with $O \left( \frac{n^2 X}{\epsilon} \right)$ Steiner
  points so that after an
  \(
    O \left(
      \frac{n^2 X}{\epsilon} \log \left( \frac{n X}{\epsilon} \right)
    \right)
  \)
  -time preprocessing phase for a given vertex $s$, we can determine a
  $(1+\epsilon)$-approximate SDP from $s$ to any point $v$ in:
  \begin{enumerate}[(i)]

  \item $O(n)$ time if $v$ is a vertex of the terrain or a Steiner
    point, and
    
  \item $O \left( \frac{n X}{\epsilon} \right)$ time otherwise.
  
  \end{enumerate}
\end{theorem}

\begin{proof}
  We first show that the path $P''$ returned by our algorithm is a
  $(1+\epsilon)$-approximation of $P$. Path $P''$ is descending
  because any path in $G$ is a descending path in the terrain, and the
  last segment of $P''$ is descending. It follows from
  Lemma~\ref{lem:Steiner.ApproxFactor}, and from the construction of
  $P''$ that the length of $P''$ is at most that of $P'$, and hence,
  $P''$ is a $(1+\epsilon)$-approximation of $P$.

  We now prove the bound on the number of Steiner points. For each
  edge $e$ of the terrain, the number of Steiner points corresponding
  to the planes $z = j \delta$ is at most $\frac{L}{\delta} - 1$, and
  the number of Steiner points corresponding to the planes $z = h(x)$
  is at most $n-2$. So,
  \(
    |V \cap e|
    \,\le\,
    \left(\frac{L}{\delta} - 1\right) + (n-2) + 2
    \,<\,
    \frac{L}{\delta} + n
    \,=\,
    4 n \left(\frac{L}{h}\right)
         \left(\frac{1}{\epsilon}\right) \sec\theta
         \,+\, n,
  \)
  because $\delta = \frac{\epsilon h \cos\theta}{4 n}$. Let
  \[
    c
    = 
    5 n \left(\frac{L}{h}\right) 
         \left(\frac{1}{\epsilon}\right) \sec\theta
    \enspace.
  \]
  Since $\left(\frac{L}{h}\right) \left(\frac{1}{\epsilon}\right)
  \sec\theta \ge 1$, we have: $|V \cap e| < c$. Using the fact that
  the number of edges is at most $3 n$, we have:
  \(
    |V|
    \le
    3 n c
    =
    O \left(
      n^2 \left(\frac{L}{h}\right) 
         \left(\frac{1}{\epsilon}\right) \sec\theta
    \right)
    =
    O \left( \frac{n^2 X}{\epsilon} \right).
  \)
  This proves the bound on the number of Steiner points.

  It follows from the running time of the Bushwhack algorithm
  (discussed in Sect.~\ref{L3:Bushwhack}) that the preprocessing time
  of our algorithm is:
  \[
    O( |V| \log |V| )
    =
    O \left(
      \frac{n^2 X}{\epsilon} \log \left(\frac{n X}{\epsilon}\right)
    \right)
    \enspace.
  \]

  During the query phase, if $v$ is a vertex of the terrain or a
  Steiner point, the approximate path is in the tree $T$. Because the
  tree has height $O(n)$, it takes $O(n)$ time to trace the
  path. Otherwise, $v$ is an interior point of a face or an edge of
  the terrain. The last intermediate node $u$ on the path to $v$ is a
  vertex or a Steiner point that lies on the boundary of a face
  containing $v$. If $v$ is interior to a face [an edge], there are 3
  [respectively 4] edges of the terrain on which $u$ can lie. Thus
  there are $O(c)$ choices for $u$, and to find the best approximate
  path we need
  \(
    O(c + n)
    = 
    O\left(
      n \left(\frac{L}{h}\right) 
          \left(\frac{1}{\epsilon}\right) \sec\theta
    \right)
    = 
    O\left( \frac{n X}{\epsilon} \right)
  \)
  time.
\end{proof}

Note that the space requirement of our algorithm is $O(|V|) = O \left(
\frac{n^2 X}{\epsilon} \right)$ since we are not storing $E$
explicitly.
  
Also note that using Dijkstra's algorithm with a Fibonacci
heap~\cite{Fredman.87} instead of the Bushwhack algorithm yields an
even simpler algorithm with a preprocessing time of
\[
  O( |V| \log |V| + |E| )
  =
  O \left(
    n^3 \left(\frac{X}{\epsilon}\right)^2
  \right)
  \enspace.
\]

When $v$ is neither a vertex of the terrain nor a Steiner point, the
query phase can be made faster by using a point location data
structure on each face. But it can be shown that the Voronoi diagram
on each face consists of hyperbolic arcs, which makes this approach
complicated.

\section{Future Work}
\label{L2:Conclusion}

We are currently working on an approximation algorithm that has less
dependence on the geometric parameters of the terrain. Although the
dependence on these parameters is natural, both the parameters
$\frac{L}{h}$ and $\sec\theta$ appear in quadratic form in the running
time of our algorithm. Moreover, they appear as a \emph{factor\/} of
$n^2$ in both the space requirement and the time requirement. Both
these points make our algorithm inefficient for a terrain with very
thin triangular faces and/or faces arbitrarily close to a horizontal
plane. Our goal is to devise an algorithm with linear or even
logarithmic dependence on $\frac{L}{h}$ and $\sec\theta$.

We are also investigating a possible direction for an exact solution
using an approach similar to Chen and Han~\cite{Chen.96}. Like other
approaches for shortest paths on polyhedral surfaces, the approach of
Chen and Han depends heavily on the fact that a (locally) shortest
path unfolds to a straight line.  To adapt their approach for our
problem, we need to solve two subproblems: extending an SDP into a new
face, and computing an SDP through a \emph{given\/} sequence of
faces. We have already derived a full characterization of the bend
angles of an SDP (more precisely, of a \emph{locally shortest
descending path\/}), which allows us to extend any such path into a
new face, and thus reduces the problem of finding an SDP from $s$ to
$t$ on a terrain to the problem of finding an SDP through a given
sequence of faces. We hope that this result will lead us to an exact
algorithm for the problem.

\end{document}